\documentclass[conference]{IEEEtran}

\usepackage{amsmath,amssymb,amsfonts}
\usepackage{cases}
\usepackage{algorithmic}
\usepackage{graphicx}
\usepackage{subfig}
\usepackage{textcomp}
\usepackage{xcolor}
\usepackage{float}
\usepackage{extarrows}

\usepackage{etoolbox}
\patchcmd{\thebibliography}{\advance\leftmargin\labelsep}{\advance\leftmargin\labelsep\itemsep-1em }{}{}

\begin{document}

\title{A robust way to speed up consensus via adaptive social networks}

\author{\IEEEauthorblockN{1\textsuperscript{st} Xunlong Wang}
\IEEEauthorblockA{\textit{School of Science} \\
\textit{Beijing University of Posts and Telecommunications}\\
Beijing, China \\
wxl\_2908@bupt.edu.cn}
\and
\IEEEauthorblockN{2\textsuperscript{nd} Bin Wu}
\IEEEauthorblockA{\textit{School of Science} \\
\textit{Beijing University of Posts and Telecommunications}\\
Beijing, China \\
bin.wu@bupt.edu.cn}
}

\maketitle
\thispagestyle{plain}

\begin{abstract}
Opinion dynamics is crucial for unraveling the complexities of human interaction in the information age. How to speed up consensus without disturbing the fate of the system is key for opinion dynamics. We propose a voter model on adaptive networks, which resembles the coevolutionary process between opinions and social relationships. We prove the existence of a one-dimensional stable manifold for the system, which facilitates us to study both the fate of the system and the consensus time it takes. Surprisingly, we find the adjustment of social relations speeds up consensus but does not affect the fate of the system.
For echo-chamber-like networks which consist of two homogeneous subnetworks connected by few sparse links, a small probability of adaptive edge dynamics is sufficient to accelerate consensus formation, which is counterintuitive. If the network structure makes consensus much slower than that of the regular networks, minor random rewiring makes a discontinuous drop in consensus time.
Our work opens up an avenue for speeding up consensus without disturbing the fate of the system. It can be insightful for crowd control.
\end{abstract}


\section{Introduction}
    Opinion dynamics play a pivotal role in shaping the way individuals interact and form collective decisions within society. This field holds immense societal importance, as it offers insights into the propagation of information and opinions \cite{noorazar2020recent}.
    It is also applied in unmanned aerial vehicles and collective intelligence \cite{bonabeau1999swarm}.

    Opinion dynamics on networks has been intensively studied \cite{gastner2019voter,sood2005voterheterogeneous}. In deterministic opinion dynamics, social phenomenon such as consensus is well studied by various dynamical models \cite{degroot1974reaching,friedkin1990social,rainer2002opinion,dandekar2013biased}. In stochastic opinion dynamics, the voter model \cite{liggett1999voter} is one of the classic models, in which individuals are more likely to adopt popular opinions. This assumption is also essential in other opinion dynamics \cite{gleeson2013binary}. For all the connected networks, consensus is inevitable. Furthermore, the invasion probability of a single novel opinion is the same for all the connected networks, provided that they share the same number of nodes \cite{tan2014towards}.
    However, topological structures, typically captured by the degree distribution, can make the consensus time extremely different \cite{tkadlec2019population}.

    Opinion polarization is a social phenomenon where individuals are surrounded by those with similar opinions and individuals have no exposure to opposing opinions.
    For the network structure, opinion polarization suggests several disconnected subnetworks with different opinions. It is ubiquitous, ranging from religion \cite{perry2022american}, race \cite{montalvo2005ethnic}, climate change \cite{mccright2011climate} to political ideology \cite{mccoy2018polarization}. The rationale behind polarization can be of two folds. On one hand, individuals stick to their own opinions \cite{plous1993psychology}. On the other hand, individuals are more likely to interact with the like-minded \cite{nickerson1998confirmation}. Polarization hinders opinion propagation between different subnetworks, thus it is detrimental to consensus.

    How to reduce opinion polarization among the population, i.e., depolarization \cite{vinokur1978depolarization,ojer2023modeling}, is crucial for reducing the political division in social science. Besides, how to speed up consensus is also significant, which is crucial for decision-making. A faster consensus formation allows groups or societies to reach decisions more quickly in time-sensitive situations, such as crisis management. However, accelerating the opinion propagation can be at the cost of reducing the consensus probability \cite{tkadlec2019population}. So a robust and fair way to speed up consensus is necessary. Here, we ask a more ambitious question: how to speed up consensus without changing the consensus probability?

    In the last two decades, edge dynamics has been introduced \cite{wu2010evolution,wu2020CCC}. The voter model on adaptive networks has been intensively studied \cite{kimura2008coevolutionary,durrett2012graph,simplex2019}.
    The edge dynamics change the network structure which can hinder opinion propagation and consensus formation. So adaptive networks may contribute to reaching consensus and even accelerating consensus, which provides a new idea to solve the problems.
    To this end, we try to figure out a robust way to speed up consensus via adaptive social networks.






\section{Model}

We propose a co-evolutionary network model incorporating opinion dynamics on the network and dynamics of the network structure.

    Let us consider a network that consists of $N$ nodes representing individuals
    and $L$ edges representing social ties. Each node holds either opinion $A$ or opinion $B$.
    The types of edges are of three types, which are $AA$, $AB$, and $BB$. We denote the average degree of the network as $\bar{k}=2L/N$. Initially, the network is connected.

    In each time step, the evolution of network structure (Process $1$) occurs with probability $1-\phi$, and the evolution of opinions (Process $2$) occurs with probability $\phi$. In this paper, we study the adaptive voter model, i.e., $0<\phi<1$.

    Let us consider the dynamics of the network structure (Process $1$).
    Firstly, an edge is randomly selected.
    Secondly, the selected edge breaks off.
    Thirdly, one of the two ends of the broken edge is selected randomly and rewires to another node at random that is not in its neighborhood to avoid multiple edges. In other words, one edge breaks off and another emerges.

    Let us consider the opinion dynamics on the network  (Process $2$).
    We adopt the voter model: a node $i$ is selected randomly. Nothing happens if node $i$ has no neighbors and Process $2$ finishes. Otherwise, a node $j$ is selected randomly in node $i's$ neighborhood. Node $i$ adopts the opinion of node $j$. Process $2$ is equivalent to a death-birth process in the structured network under neutral selection \cite{traulsen2009stochastic}.

    Isolated network structures
    appear from time to time during the rewiring
    process.
    But isolated sub-networks will be connected via the rewire-to-random property of Process $1$. Thus, opinions spread to every corner of the network sooner or later.
    In other words, consensus (i.e., All $A$ or All $B$) is the only absorbing state.

\section{Results}
    We are interested in how the entire network reaches consensus. In this section, we show i) the most likely trajectory via which the system reaches consensus; ii) the probability of consensus of a given opinion; and iii) the time it takes to reach consensus.

    $N$, $L$, $\bar{k}$ are constants over the coevolutionary process because neither Process $1$ nor Process $2$ changes the total number of nodes and the total number of edges. Nevertheless, the proportion of opinion and the proportion of edges of types are changing.

    We denote the number of nodes holding opinion $X$ by $[X]$ and the number of $X-Y$ edges by $[XY]$, where $X, Y \in \{A, B\}$ and $[XY]=[YX]$. The system is captured by the following five variables $[AA]$, $[BB]$, $[AB]$, $[A]$, and $[B]$. Triplets (i.e. $X-Y-Z$, where the central $Y$ has $X-Y$ and $Y-Z$ edges) are introduced. Denote the number of $X-Y-Z$ triplets by $[XYZ]$. Due to the symmetry of triplets, we have that $[XYZ]=[ZYX]$.

    For $[AA]$, both Process $1$ and Process $2$ can change it. Let us first consider how Process $1$ influences $[AA]$. If an $A-B$ edge is selected and breaks off, and if the node holding opinion $A$ at the ends of the broken edge rewires to a node holding opinion $A$ in the network, then $[AA]$ increases by one. If an $A-A$ edge is selected and breaks off, the selected node of the broken edge has to be $A$, and the selected node rewires to another node in the network that holds opinion $B$, then $[AA]$ decreases by one. Let us secondly consider how Process $1$ influences $[AA]$. If a node holding opinion $B$ is selected, and if a node in its neighborhood who holds opinion $A$ is selected too, and if the former node adopts the opinion of the latter one, then $[AA]$ increases by one. If a node with opinion $B$, which has two neighbors holding opinion $A$, is selected (i.e.,  the node in the center of a triplet $A-B-A$ is selected), and if the center node adopts the opinion of any neighbor holding opinion $A$, then $[AA]$ increases by two. If a node with opinion $A$, which has one neighbor with opinion $A$ and another neighbor with opinion $B$, is selected (i.e., the node in the center of a triplet $A-A-B$ is selected), and if the center node adopts the opinion of the neighbor holding opinion $B$, then $[AA]$ decreases by one. Therefore, we obtain the mean-field equation of $[AA]$ based on the co-evolutionary process,
    \begin{align}
        \frac{d[AA]}{dt}\!=&\!\left\{\! \frac{[B]}{N}R(A|B)\!+\! 2\frac{[B]}{N}R(ABA|B)\!-\!\frac{[A]}{N}R(AAB|A)\!\right\} \nonumber\\
        &\times\!\phi\!+\!\left\{ \frac{[AB]}{L}\cdot\frac{1}{2}\frac{[A]}{N}\!-\!\frac{[AA]}{L}\cdot \frac{[B]}{N}\right\}\!\times\!(1\!-\!\phi).\label{eq:1}
    \end{align}

    Given a node with opinion $X$ is selected, $R(Y|X)$ denotes the rate at which the node with opinion $X$ finds a neighbor with opinion $Y$;
    $R(YXZ|X)$ denotes the rate at which the node with opinion $X$ randomly chooses two neighbors, one of which is with opinion $Y$ and the other of which is with opinion $Z$ (i.e., the selected node is in the center of a triplet $Y-X-Z$).

    Similarly, we write the mean-field equations of $[BB]$ and $[A]$ based on the co-evolutionary process,
    \begin{align}
        \frac{d[BB]}{dt}\!=&\!\left\{\! \frac{[A]}{N}R(B|A)\!+\! 2\frac{[A]}{N}R(BAB|A)\!-\!\frac{[B]}{N}R(ABB|A)\!\right\} \nonumber\\
        &\times\!\phi\!+\!\left\{\! \frac{[AB]}{L}\cdot\frac{1}{2}\frac{[B]}{N}\!-\!\frac{[BB]}{L}\cdot\frac{[A]}{N}\!\right\}\!\times\!(1\!-\!\phi),\label{eq:2}\\
        \frac{d[A]}{dt}=&\phi\times\left\{\frac{[B]}{N}\cdot R(A|B)-\frac{[A]}{N}\cdot R(B|A)\right\}.\label{eq:3}
    \end{align}

    We use the pair-approximation to simplify the system. In other words, we assume that $R(Y|X)=[XY]/(\sum_{W\neq X}[XW]+2[XX])$ and
    $R(YXZ|X)=[YXZ]/(\sum_{W\neq X}[XW]+2[XX])$ where $[XYX]=[XY][XY]/(2[Y])$ and $[XXY]=2[XX][XY]/[X]$ \cite{kimura2008coevolutionary}.

    Note that the total number of nodes and the total number of edges in the network remain constant (i.e., $[AA]+[BB]+[AB]\equiv L$ and $[A]+[B]\equiv N$), we only need to study three variables $[AA]$, $[BB]$, and $[A]$. Thus, \eqref{eq:1}-\eqref{eq:3} are closed.

    We take a coordination transformation. Setting $u=([AA]+[BB])/L$, $v=([AA]-[BB])/L$, and $w=([A]-[B])/N$, \eqref{eq:1}-\eqref{eq:3} become
    \begin{numcases}{}
        \frac{du}{dt}=\phi\frac{2}{N\bar{k}}\cdot\frac{1-u}{1-v^2}\left[1-vw+\bar{k}(1-2u+v^2)\right]\nonumber\\\qquad\ +\frac{1-\phi}{N\bar{k}}\cdot (1-2u+vw),\label{eq:4}\\
        \frac{dv}{dt}=\left[\phi\frac{2}{N\bar{k}}\cdot\frac{1-u}{1-v^2}-\frac{1-\phi}{N\bar{k}}\right](v-w),\label{eq:5}\\
        \frac{dw}{dt}=\phi\frac{2}{N}\cdot\frac{1-u}{1-v^2}\cdot(v-w),\label{eq:6}
    \end{numcases}
    where $0\le u\le 1$, $-1\le v\le 1$ and $-1\le w\le 1$.

    $u$ represents the proportion of homogeneous edges (i.e., both ends of the edge hold the same opinion).
    If $u>1/2$, then more than half of the social ties are between those with the same opinion.
    $w$ represents the popularity of opinion $A$. When $w$ is positive, opinion $A$ is more popular than opinion $B$. When $w$ is equal to $1$, every node in the population holds opinion $A$ (i.e., All $A$).
    Every node holds opinion $B$ (i.e., All $B$) when $w$ is equal to $-1$.

\subsection{Stable Manifold}
    We find that a set of fixed points emerges in \eqref{eq:4}-\eqref{eq:6}, which is given by
    \begin{numcases}{}
        u = 1-\alpha(1-w^2),\label{eq:7}\\
        v = w,\label{eq:8}
    \end{numcases}
    where $\alpha=\frac{\phi(\bar{k}-1)-(1-\phi)+\sqrt{(1-\phi)^2+2(1+\bar{k})\phi(1-\phi)+\phi^2(\bar{k}-1)^2}}{4\phi\bar{k}}$  has nothing to do with the three variables, and only depends on the average degree $\bar{k}$ and the probability of Process $2$ occurring $\phi$.

    Next, we prove the set of fixed points is a one-dimensional stable manifold.

    Take a point on the set of fixed points. Denote it as $p_1$, where $p_1=[u_1,v_1,w_1]^T=[1-\alpha(1-{w_1}^2),w_1,w_1]^T$. Linearizing the system at point $p_1$, we obtain
    \begin{equation}\label{eq:9}
    \frac{d[u,v,w]^T}{dt}\Bigg|_{p_1}\approx
    \left[
    \begin{array}{ccc}
        d & aw_1 & bw_1\\
        0 & -b & b\\
        0 & c & -c
    \end{array}
    \right]\cdot
    \left[
    \begin{array}{c}
         u \\
         v\\
         w
    \end{array}
    \right],
    \end{equation}
    whose correction term is of $O(u^2+v^2+w^2)$.

    Here,
    \begin{numcases}{}
        a=\frac{2\phi\alpha(1+4\bar{k}\alpha)+(1-\phi)}{N\bar{k}},\\
        b=\frac{1-\phi-2\phi\alpha}{N\bar{k}},\\
        c=\frac{2\phi\alpha}{N},\\
        d=-\frac{2\sqrt{(1-\phi)^2+2(1+\bar{k})\phi(1-\phi)+\phi^2(\bar{k}-1)^2}}{N\bar{k}},
    \end{numcases}
     where $a$, $b$ and $c$ are all positive and $d$ is negative when $\bar{k}\ge 2$. It must be a disconnected network for $\bar{k}<2$, so we do not consider $\bar{k}<2$.

     The eigenvalues at the set of fixed points \eqref{eq:7}-\eqref{eq:8} are
    \begin{align}
        \lambda=
        \left\{
        \begin{array}{l}
             0 \ , \\
             d \ ,\\
             -b-c \ .
        \end{array}\label{eq:14}
        \right.
    \end{align}

    Since only one of the eigenvalues \eqref{eq:14} is $0$ and the rest are negative, we find that the set of fixed points \eqref{eq:7}-\eqref{eq:8} is a one-dimensional stable manifold based on the center manifold theorem \cite{2002Nonlinear,strogatz2018nonlinear}.
    The stability of the manifold implies that the system converges to the quadratic curve depicted by \eqref{eq:7}-\eqref{eq:8}, no matter what the initial network configuration and the popularity of opinions are.

\subsection{Consensus Probability}

    The stable manifold further facilitates us to capture the co-evolutionary process as a one-dimensional Markov chain with state variable $w$, or equivalently the number of nodes with opinion $A$ (i.e., $[A]$). Interestingly, $[A]$ is directly changed only by Process $2$, not Process $1$.

    The state space of the one-dimensional Markov chain is thus $\Omega=\{0,1,2,\cdots, N\}$. Each state represents the number of nodes with opinion $A$. Herein both $0$ (All $B$) and $N$ (All $A$) are the absorbing states of the Markov chain, and the fate of the system is determined by which absorbing state it gets absorbed into.
    The transition probability of increasing by one from state $j$, $P^{+}_j$, is given by
    \begin{align}
        P^+_j&=\phi\cdot\frac{[B]}{N}\cdot R(A|B)\nonumber\\
            &=\phi\cdot\frac{[B]}{N}\cdot\frac{[AB]}{[AB]+2[BB]}\nonumber\\
            &=\phi\cdot\frac{1-w}{2}\cdot\frac{1-u}{1-v}\bigg|_{w=\frac{2j}{N}-1}\nonumber\\
            &\xlongequal[v=w]{u=1-\alpha(1-w^2)} \frac{\alpha\phi}{2}(1-w^2)\bigg|_{w=\frac{2j}{N}-1}\nonumber\\
            &= \frac{\alpha\phi}{2}[1-(\frac{2j}{N}-1)^2]\label{eq:15}
    \end{align}
    Similarly, The transition probability of decreasing by one from state $j$, $P^{-}_j$, is obtained.
    \begin{align}
        P^-_j=\frac{\alpha\phi}{2}[1-(\frac{2j}{N}-1)^2]\label{eq:16}
    \end{align}

    To capture the fate of the system, we are interested in the likelihood that $j$ opinion $A$ nodes take over the system. Consensus probability $\Phi_j$ denotes the probability that the Markov chain starts from state $j$ and is eventually absorbed by state $N$. In other words, consensus probability is the probability that the fate of the system is All $A$.
    We find $P^+_j$ is equal to $P^-_j$, which implies that the motion in the one-dimensional manifold is a one-dimensional diffusion process with neutral drift \cite{traulsen2009stochastic,gardiner2012handbook}, where $\Phi_j=j/N$.
    It's counterintuitive that the fate of the system has nothing to do with both the adjustment of social relations and the average degree. In other words, the adjustment of social relations does not affect the final consensus outcome. Besides, the consensus probability of the adaptive network is equivalent to that of a static homogeneous network \cite{sui2015speed,maciejewski2014reproductive}, which implies the introduction of adaptive edge dynamics doesn't alter the fate of the system.

    \subsection{Consensus Time}

    Consensus time $T_j$ denotes the average time that the Markov chain starts from state $j$ and is eventually absorbed by state $0$ or state $N$. In other words, consensus time is the time taken to reach a consensus, where it could be either All $A$ or All $B$.
    We obtain the consensus time by \cite{traulsen2009stochastic,gardiner2012handbook},
    \begin{align}
        T_1=&\frac{1}{N}\sum_{k=1}^{N-1}\sum_{l=1}^k\frac{1}{P^+_l}\nonumber\\
        =&\frac{1}{N}\sum_{k=1}^{N-1}\sum_{l=1}^k\frac{1}{\alpha\phi}\cdot\frac{2}{1-w^2}\bigg|_{w=\frac{2j}{N}-1}\nonumber\\
        =&\frac{1}{N}\sum_{k=1}^{N-1}\sum_{l=1}^k\frac{1}{\alpha\phi}\left(\frac{1}{1-w}+\frac{1}{1+w}\right)\bigg|_{w=\frac{2j}{N}-1}\nonumber\\
        =&\frac{1}{N}\sum_{k=1}^{N-1}\sum_{l=1}^k\frac{N}{2\alpha\phi}\left(\frac{1}{N-l}+\frac{1}{l}\right)\nonumber\\
        =&\frac{1}{2\alpha\phi}\sum_{k=1}^{N-1}\sum_{l=1}^k\left(\frac{1}{N-l}+\frac{1}{l}\right).\label{eq:19}
    \end{align}

    The sums in the previous equation \eqref{eq:19} can be interpreted as numerical approximations to the integrals (i.e., $\sum_{k=1}^i\ldots\approx\int_1^i\ldots dk$) \cite{traulsen2006stochastic,traulsen2009stochastic}. Replacing the sums with the integrals, we obtain
   \begin{align}
       T_1\approx&\frac{1}{2\alpha\phi}\int_1^{N-1}\left[\int_1^k \left(\frac{1}{N-l}+\frac{1}{l}\right)dl\right]dk\nonumber\\
       =&\frac{1}{2\alpha\phi}(N-1)\ln(N-1)\nonumber\\
       \approx&\frac{1}{2\alpha\phi}N\ln(N-1).\label{eq:20}
   \end{align}

   We find that if $\phi=1$, $T_1$ basically agrees with the fixation time on regular graphs under neutral selection \cite{sui2015speed}.
   Similarly, $T_j$ is approximated by

    \begin{align}
        \frac{N}{2\alpha\phi}\left\{(N\!-\!1)\ln(N\!-\!1)\!-\![(N\!-\!j)\ln(N\!-\!j)\!+\!j\ln j]\right\}.\label{eq:21}
    \end{align}

    Based on \eqref{eq:21}, we find that if $j=N/2$ (Assuming $N$ is an even number), the consensus time $T_j$ reaches its maximum. It is given by
    \begin{align}
        T_{\frac{N}{2}}&=\frac{1}{2\alpha\phi}\left[(N^2-N)\ln(N-1)-N^2\ln\frac{N}{2}\right]\label{eq:22}\\
        &=\frac{1}{2\alpha\phi}\left[N^2\ln\frac{2N-2}{N}-N\ln (N-1)\right]\nonumber\\
        &=\frac{N^2}{2\alpha\phi}\left[\ln(2-\frac{2}{N})-\frac{\ln(N-1)}{N}\right]\nonumber\\
        &\xrightarrow{N\to \infty}\frac{\ln2}{2\alpha\phi}N^2,\label{eq:23}
    \end{align}
   which implies that the maximum consensus time is of $O(N^2)$.

    Opinion propagation depends only on Process $2$ rather than Process $1$. So we are also interested in the number of occurrences of Process $2$ before consensus, which indicates the impact of opinion propagation on consensus. We call it the pure consensus time, denoted by $\hat{T_j}$, where $\hat{T_j}=T_j\phi$. When $\hat{T_j}$ is small, it indicates that the system reaches consensus with a small number of opinion propagation. In other words, the smaller $\hat{T_j}$ is, the stronger the impact of opinion propagation on consensus is.

    We find that $T_j$ decreases with $\phi$, which is intuitive because the speedup of opinion propagation contributes to consensus formation. However, it is counterintuitive that $\hat{T_j}$ increases with $\phi$. It implies that the adjustment of the network structure enhances the impact of opinion propagation on consensus. Therefore, the adjustment of social relations enhances the impact of opinion propagation on consensus but does not affect the consensus outcome.

\section{Depolarization via adaptive networks}
    \begin{figure}[htbp]
        \centering
        \includegraphics[width=0.5\textwidth]{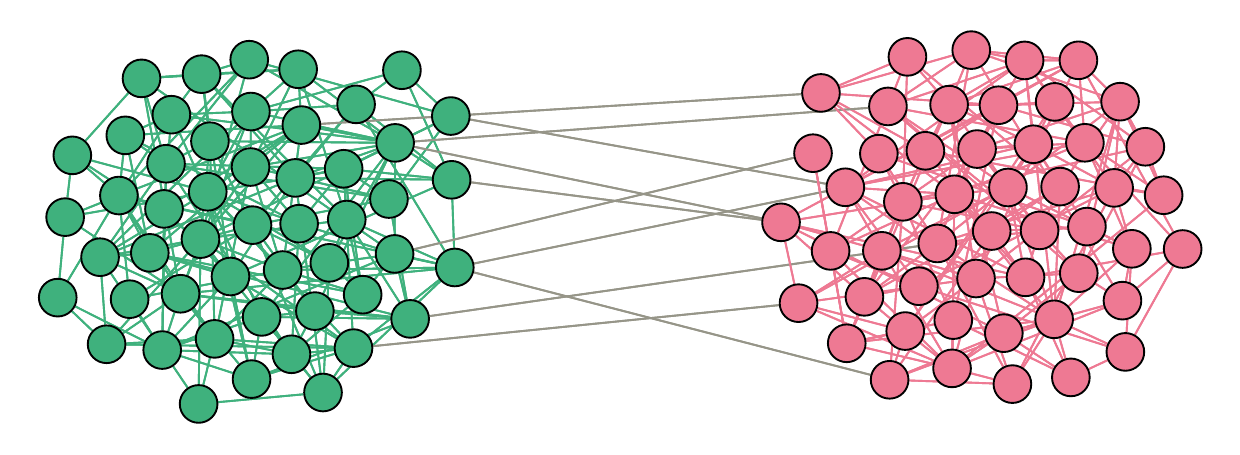}
        \caption{\textbf{The echo-chamber-like networks have a long consensus time.} An echo-chamber-like network consists of two homogeneous subnetworks with different opinions connected by few sparse links, which largely hinder opinion propagation and slow down consensus. Here is an example. The network consisting of $100$ nodes and $400$ edges has only $10$ $A-B$ edges. The red nodes represent nodes with opinion $A$ and the green nodes represent nodes with opinion $B$.}
        \label{fig: echo}
    \end{figure}

    Individuals are more likely to interact with like-minded partners. Thus people are likely to be surrounded by individuals who share the same opinions. And individuals frequently lack exposure to information presenting opposing opinions, which is referred to as echo-chambers \cite{wang2020public,baumann2020modeling}. For the evolution of the network structure, echo-chamber effects generate echo-chamber-like networks, which consist of two homogeneous subnetworks connected by few sparse links. The two subnetworks represent two different opinions.

    \begin{figure*}[htbp]
        \centering
        \subfloat[$\phi=0.2$]{\includegraphics[width=0.33\textwidth]{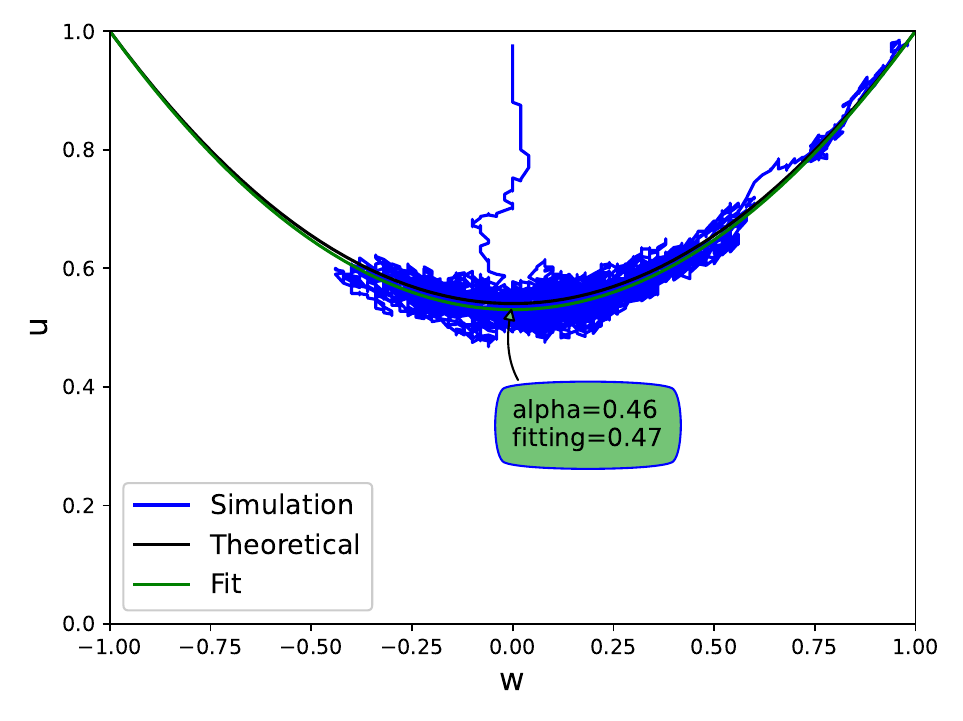}\label{fig: 20}}
        \hfill
        \subfloat[$\phi=0.5$]{\includegraphics[width=0.33\textwidth]{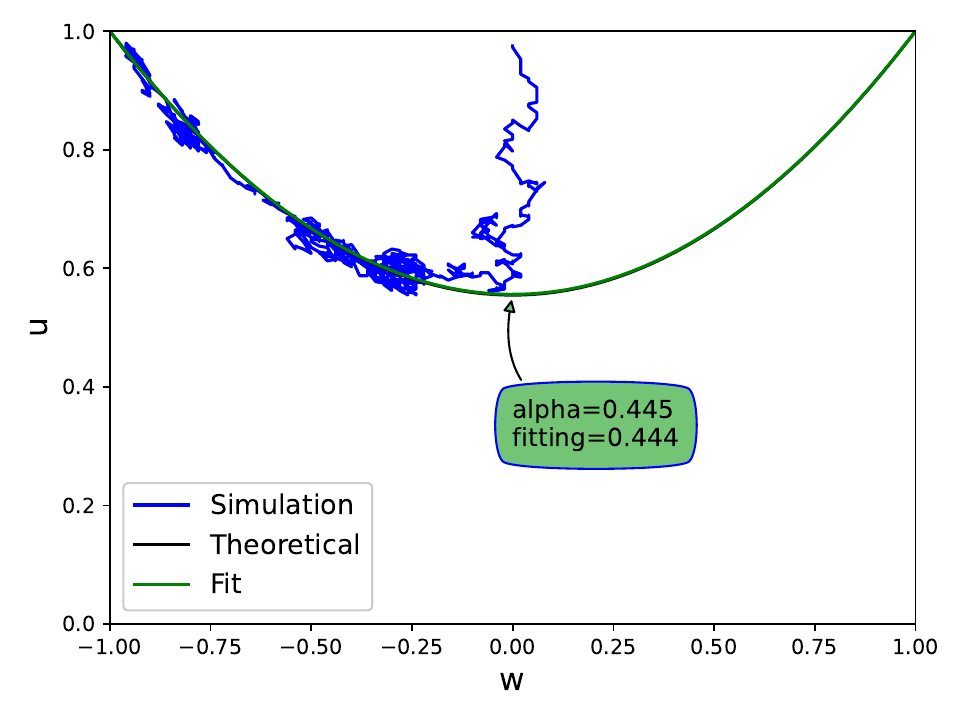}\label{fig: 50}}
        \hfill
        \subfloat[$\phi=0.8$]{\includegraphics[width=0.33\textwidth]{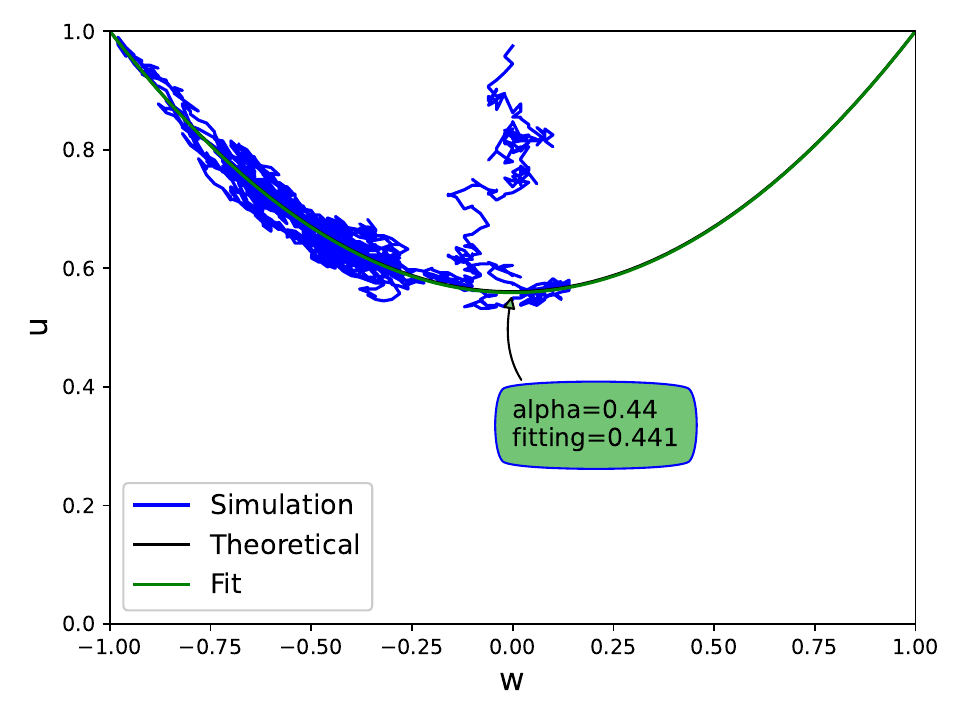}\label{fig: 80}}
        \caption{\textbf{The most likely trajectory is a quadratic curve.} We show sample paths of the dynamics in the $(w,u)$-space at voting probabilities of (a) $0.2$; (b) $0.5$; (c) $0.8$. The points in the $(w,u)$-space quickly converge to the stable manifold and follow it to the consensus. The $alpha$ in the green box is the $\alpha$ in \eqref{eq:7}, where $\alpha=$ (a) $0.46$; (b) $0.445$; (c) $0.44$. The green line is the best fit of the path to the parameterized parabola $u=1-\xi(1-w^2)$. In order to avoid the effect of the pre-convergence trajectory, we fit with the last ninety percent of the trajectory. The best fitting value of $\xi$ is presented as $fitting$ in the green box and $\xi=$ (a) $0.47$; (b) $0.444$; (c) $0.441$. The green parabola almost coincides with the black one, which confirms the accuracy of the theoretical solution.}
        \label{fig: manifold}
    \end{figure*}

    We take Fig. \ref{fig: echo} as the initial network and conduct multiple sets of simulations.
    As shown in Fig. \ref{fig: manifold}, the trajectories initially at $(w,u)=(0,1)$ converge quickly to the stable manifold. The simulation result confirms the existence and the stability of the one-dimensional manifold. Furthermore, it also indicates the accuracy of the theoretical solutions \eqref{eq:7}\eqref{eq:8}. Based on Fig. \ref{fig: manifold}, we find that as $\phi$ increases, the convergence slows down for echo-chamber-like networks. It also suggests that the adjustment of social relations accelerates the propagation of opinions.

    \begin{figure*}[htbp]
        \centering
        \subfloat[Consensus Probability]{\includegraphics[width=0.495\textwidth,height=6.5cm]{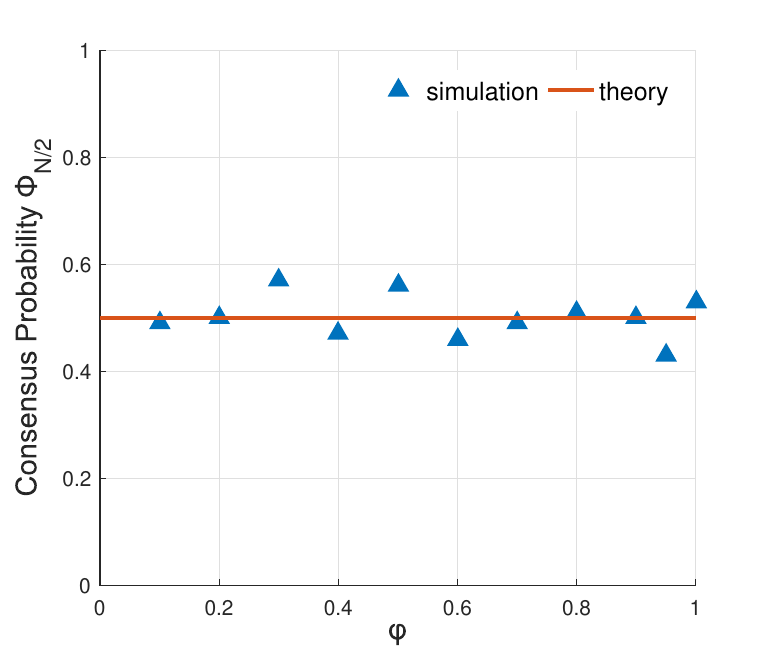}\label{fig: prob}}
        \hfill
        \subfloat[Consensus Time]{\includegraphics[width=0.495\textwidth,height=6.5cm]{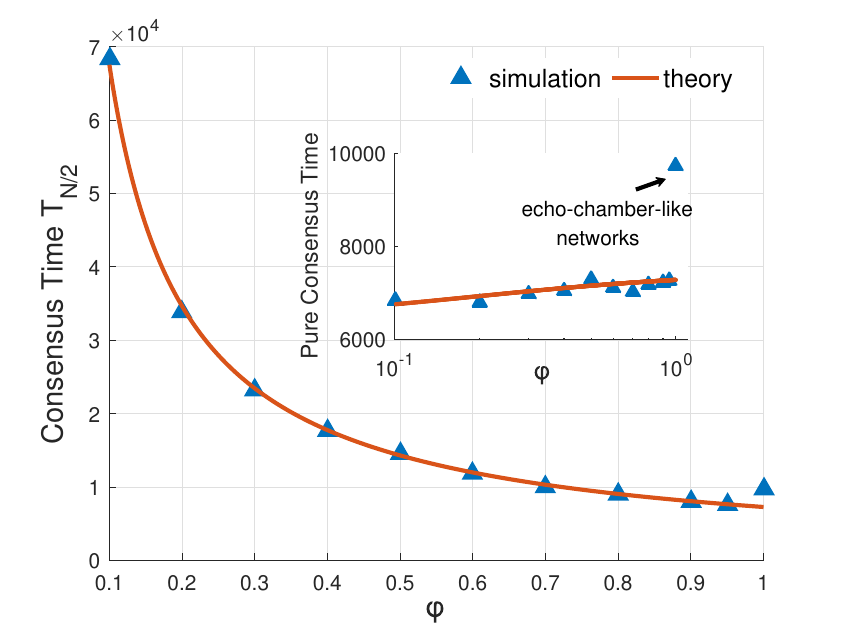}\label{fig: time}}
        \hfill
        \caption{\textbf{Random rewiring is an effective way to speed up consensus without changing the fixation probability.} As the adjustment of the network structure increases, i.e., as $\phi$ decreases, the consensus probability is invariant, whereas the consensus time increases. However, the pure consensus time increases with $\phi$. If the initial network is a static echo-chamber-like network, the consensus time is much longer than others. But only $5\%$ edge dynamics is sufficient to reduce the consensus time to less than $80\%$. The red lines represent the analytical results, and the triangles are the simulation results. For every realization, all the simulation data points are calculated by averaging over 100 independent runs.}
        \label{fig: final}
    \end{figure*}

    As shown in Fig. \ref{fig: final}, the simulation results validate our theoretical solutions of the consensus probability, the consensus time, and the pure consensus time. Based on Fig. \ref{fig: time}, the pure consensus time increases slowly with $\phi$ rises, which implies that the adjustment of social relations enhances the impact of opinion propagation on consensus. Surprisingly, we find a difference in pure consensus time between $\phi=0.95$ and $\phi=1$. It implies that for echo-chamber-like networks, the introduction of even a small probability of adaptive edge dynamics significantly accelerates consensus formation, which is counterintuitive.
    For the network that hinders opinion propagation, minor random rewiring makes a discontinuous drop in the consensus time and makes it converge to that of static regular networks.
    To sum up, the results in Fig. \ref{fig: final} validate that it's a robust way to speed up consensus.

\section{Discussion \& Conclusion}

    Speeding up consensus is vital for improving the efficiency of decision-making processes, resolving conflicts, and promoting social cohesion. However, a robust way to accelerate consensus formation is still lacking.

    We find that for echo-chamber-like networks, the introduction of even a small probability of adaptive edge dynamics is sufficient to effectively speed up consensus.
    Only $5\%$ of the edge dynamics shrink the consensus time to less than $80\%$, which is counterintuitive and validates the discontinuous drop in consensus time.
    Meanwhile, the social network adjustment is so naive that there is no social bias, but the impact is significant.
    We thus make the conjecture that the introduction of a minor adjustment of the network structure makes the consensus time converge to that of the static regular networks \cite{sui2015speed}.
    The adjustment of the network structure reinforces the impact of opinion propagation on consensus and does not interfere with the final consensus outcome. Thus random rewiring is a robust way to speed up consensus.

    Our work also has provided a novel theoretical method to analytically deal with the coevolutionary dynamics.
    Analytical methods on present coevolutionary dynamics are typically based on the assumption that social interactions evolve much faster than opinions \cite{wu2010evolution,wu2011evolutionary,wu2016evolving,wu2019evolution,wu2020CCC,wei2019vaccination,shan2022social,wang2023opinions,liu2023emergence,baumann2020modeling}.
    The assumption is technically favored because the network structure has converged to a steady state whenever the opinion is updated. In other words, The steady state of the network structure is key for previous analysis of opinion dynamics.
    However, the real co-evolutionary process is not consistent with the assumption. Our analysis extrapolates from this assumption and solves the resulting technical difficulties. Our method is still valid if the social relation dynamics and opinion dynamics are evolving on the relatively same time scale, which was not achievable by previous works.
    Furthermore, we analytically show the trajectory via which the system reaches consensus, instead of fitting parabolas \cite{simplex2019} or numerical solutions \cite{durrett2012graph,marceau2010adaptive}. The theoretical solution facilitates us to estimate both the consensus probability and the consensus time, but the solution obtained by fitting or numerical calculation cannot do it.


    Research on how to speed up opinion propagation without disturbing the final outcome is of significance for opinion dynamics. It is insightful for the control of public opinion and the spread of information.
    We show that it's a robust way to speed up consensus via adaptive social networks.

\section*{Acknowledgment}
    We gratefully acknowledge Yakun Wang, who helps us polish figures. We appreciate the sponsorship from NSFC No.61751301.
\bibliographystyle{unsrt}
\bibliography{output}
\end{document}